\documentclass[twocolumn,11pt]{article}
\usepackage{times}
\usepackage{graphicx}
\usepackage{amsmath}
\usepackage{amsfonts}
\usepackage{xcolor}
\usepackage{url}
\begin{document}
\global\def\refname{{\normalsize \it References:}}
\baselineskip 12.5pt
%
%
%
\title{\LARGE \bf Clifford Algebra with \emph{Mathematica}}

\date{}

\author{\hspace*{-10pt}
\begin{minipage}[t]{2.7in} \normalsize \baselineskip 12.5pt
\centerline{J.L. ARAG\'{O}N}
\centerline{Universidad Nacional Aut\'{o}noma de M\'{e}xico}
\centerline{Centro de F\'{\i}sica Aplicada}
\centerline{y Tecnolog\'{\i}a Avanzada}
\centerline{Apartado Postal 1-1010, 76000 Quer\'{e}taro}
\centerline{MEXICO}
\centerline{aragon@fata.unam.mx}
\end{minipage} \kern 0in
\begin{minipage}[t]{2.7in} \normalsize \baselineskip 12.5pt
\centerline{G. ARAGON-CAMARASA}
\centerline{University of Glasgow}
\centerline{School of Computing Science}
\centerline{Sir Alwyn William Building,}
\centerline{Glasgow, G12 8QQ Scotland}
\centerline{UNITED KINGDOM}
\centerline{Gerardo.AragonCamarasa@glasgow.ac.uk}
\end{minipage} \\ ~ \\
\begin{minipage}[t]{2.7in} \normalsize \baselineskip 12.5pt
\centerline{G. ARAG\'{O}N-GONZ\'{A}LEZ}
\centerline{Universidad Aut\'{o}noma Metropolitana}
\centerline{Unidad Azcapotzalco}
\centerline{San Pablo 180, Colonia
	Reynosa-Tamaulipas,}
\centerline{02200 D.F. M\'{e}xico}
\centerline{MEXICO}
\centerline{gag@correo.azc.uam.mx}
\end{minipage}
\begin{minipage}[t]{2.7in} \normalsize \baselineskip 12.5pt
\centerline{M.A. RODR\'{I}GUEZ-ANDRADE}
\centerline{Instituto Polit\'{e}cnico Nacional}
\centerline{Departamento de Matem\'{a}ticas, ESFM}
\centerline{UP Adolfo L\'{o}pez Mateos,}
\centerline{Edificio 9. 07300 D.F. M\'{e}xico}
\centerline{MEXICO}
\centerline{marco@polaris.esfm.ipn.mx}
\end{minipage}
%
%
\\ \\ \hspace*{-10pt}
\begin{minipage}[b]{6.9in} \normalsize
\baselineskip 12.5pt {\it Abstract:}
The Clifford algebra of a n-dimensional Euclidean vector space
provides a general language comprising vectors, complex numbers,
quaternions, Grassman algebra, Pauli and Dirac matrices. In this work,
we present an introduction to the main ideas of Clifford algebra, with
the main goal to develop a package for Clifford algebra calculations
for the computer algebra program
\textit{Mathematica}\footnote{\textit{Mathematica} is a registered
  trademark of Wolfram Research, Inc.}. The Clifford algebra package
is thus a powerful tool since it allows the manipulation of all
Clifford mathematical objects. The package also provides a
visualization tool for elements of Clifford Algebra in the
3-dimensional space. \texttt{clifford.m} is available from\\
{\centering \url{https://github.com/jlaragonvera/Geometric-Algebra}}
\\ [4mm] {\it Key--Words:}
Clifford Algebras, Geometric Algebra, Mathematica Software.
\end{minipage}
\vspace{-10pt}}

\maketitle

\thispagestyle{empty} \pagestyle{empty}
%
%
\section{Introduction}
\label{S1} \vspace{-4pt}

The importance of Clifford algebra was recognized for the first time
in quantum field theory. Lately, there has been a tendency to
exploit their power in many others fields. These fields include
projective geometry \cite{HesZieg}, electrodynamics \cite{Jancewicz},
analysis on manifolds and differential geometry \cite{HesSob},
crystallography \cite{Aragon1,Aragon2} to name a few. A recent account of
the applications of Clifford algebra in fields such as robotics,
computer vision, computer graphics, engineering, neural and quantum
computing, etc., can be found in \cite{Bayro} and \cite{Sommer}.

General introductions to Clifford algebra can be found in several
books (see for instance Refs. \cite{Porteous} and
\cite{Delanghe}). Here, a gently introduction to the Clifford algebra
of $\mathbb{R}^n$ is presented together with some examples to show the
generality of this algebra in order to provide a general language
comprising vectors, complex numbers and quaternions. The main goal is
to implement the basic operations of Clifford algebras in
\textit{Mathematica}, resulting in a package for doing Clifford
algebra computations. There exists some other packages and specialized
programs for doing Clifford algebra; \emph{CLIFFORD/Bigebra} is a
\emph{Maple} package which include additional specialized packages
such as \emph{SchurFkt} (for the Hopf Algebra of Symmetric Functions)
and \emph{GfG - Groebner for Grassmann} (for Computing Groebner Bases
for Ideals in Grassmann Algebra) \cite{Ablamowicz};
\emph{TCliffordAlgebra} is as add-on application for the
\emph{Mathematica} package \emph{Tensorial} that implements Clifford
algebra operations \cite{Cabrera}; \emph{CLICAL} is a stand-alone
calculator-type computer program for \emph{MS-DOS}
\cite{Lounesto}. While the first two packages requires more
specialized knowledge of Clifford algebras, \emph{CLICAL} and the
package presented here is easy to use and can be used by non
mathematicians. More recently, \emph{Clifford Multivector Toolbox}, a
toolbox for computing with Clifford algebras in \emph{MATLAB}, has
been released \cite{Ehitzer}. A more specialized package is
\emph{GAViewer} package, designed for computation and visualization of
objects in conformal geometric algebra \cite{Dorst} (for a more recent
account and application see Ref. \cite{Kanatani}). We must emphasise
that with the exception of the \emph{Maple} packages, the other
packages and toolboxes only allow numeric computations. Our Clifford
Mathematica package allow us to carry out numerical and symbolic
computations with complex numbers, quaternions, the hyperbolic plane,
Grassmann algebra and, Dirac and Pauli algebras, all defined within
the Clifford algebra framework. Our package {\small \verb|clifford.m|}
is available for download at:

{\scriptsize \noindent \url{https://github.com/jlaragonvera/Geometric-Algebra}}

\section{The Clifford algebra of $\mathbb{R}^n$}
\label{S2} \vspace{-4pt}

The set of n-tuples of the form $(x_1 ,x_2 ,...,x_n )$ with the
standard operations of addition and multiplication for real numbers is
a vector space, over the field of real numbers, which we denote as
$\mathbb{R}^n$. It means that the addition of $n$-tuples, and
multiplication by real numbers satisfy certain properties which are
those of a vector space \cite{Lang}.  The canonical basis of
$\mathbb{R}^n$ is the set of $n$-dimensional vectors $\{
{\mathbf{e}}_1 , {\mathbf{e}}_2 ,..., {\mathbf{e}}_n \}$ where
$\langle {\mathbf{e}}_i , {\mathbf{e}}_j \rangle = \delta _{ij} $ and
$\langle \; , \; \rangle$ is a inner product in $\mathbb{R}^n$. An
element ${\mathbf{v}}$ of $\mathbb{R}^n$ is written as a linear
combination of this canonical basis:
\begin{equation*}
  v = x_1 {\mathbf{e}}_1 + x_2 {\mathbf{e}}_2 + \cdots + x_n {\mathbf{e}}_n .
\end{equation*}

It is said therefore that the n-tuple $(x_1 ,x_2 ,...,x_n )$ is the
coordinate vector of ${\mathbf{v}}$ with respect to the canonical
basis $\{ {\mathbf{e}}_1 ,{\mathbf{e}}_2 ,...,{\mathbf{e}}_n \}$.

The vector space $\mathbb{R}^n$ has two operations defined:
addition of vectors and multiplication of vectors by scalars. The
multiplication by vectors between themselves is not defined. The
algebraic structure which considers multiplication between vectors is
called an algebra.

An algebra ${\mathcal{A}}$ is a vector space over a field
${\mathcal{F}}$ together with a binary multiplication $\mathbf{ab}$ in
${\mathcal{A}}$ such that form any
${\mathbf{a}}, {\mathbf{b}}, {\mathbf{c}} \in {\mathcal{A}}$ and
$\alpha \in {\mathcal{F}}$ \cite{Jacobson}:
\begin{eqnarray*}
  ({\mathbf{a}} + {\mathbf{b}}) {\mathbf{c}} & = & \mathbf{ac} + 
                                                   \mathbf{bc} \\
  {\mathbf{a}} ( {\mathbf{b}} + {\mathbf{c}} ) & = & \mathbf{ab} + 
                                                     \mathbf{ac} \\
  \alpha ( \mathbf{ab} ) & = & (\alpha {\mathbf{a}} ) {\mathbf{b}} = 
                               {\mathbf{a}} (\alpha {\mathbf{b}} ) .
\end{eqnarray*}

In the case of the vector space $\mathbb{R}^n$, the field
${\mathcal{F}}$ is the set of real numbers. In
order to construct an algebra from $\mathbb{R}^n$, it is required to define
a product $\mathbf{ab}$ between vectors in $\mathbb{R}^n$. One
particular product in $\mathbb{R}^n$ can be defined as follows.

Let us consider the vector space $\mathbb{R}^n$ with the inner product
$\langle {\mathbf{a}}, {\mathbf{b}} \rangle$ and an orthonormal basis
$\{ {\mathbf{e}}_1 ,{\mathbf{e}}_2 ,...,{\mathbf{e}}_n \}$. We
construct an algebra from $\mathbb{R}^n$ by introducing a product
between vectors in $\mathbb{R}^n$ that satisfies the condition
\begin{equation}
 \label{product}
 \mathbf{ab} + \mathbf{ba} = 2 \langle {\mathbf{a}} , {\mathbf{b}} 
 \rangle  {\mathbf{1}},
\end{equation}
where ${\mathbf{1}}$ is the identity of the algebra. The product so
defined is associative:
\begin{equation*}
  {\mathbf{a}}(\mathbf{bc})=(\mathbf{ab}){\mathbf{c}},
\end{equation*}
and we are constructing an algebra ${\mathcal{A}}$ equipped with an
identity ${\mathbf{1}}$.

With the vector space and the product (\ref{product}), the resulting
algebra of all possible sums and products of vectors in
$\mathbb{R}^{n}$ is called the Clifford algebra of $\mathbb{R}^{n}$
and is denoted by ${\mathcal{C}l}_{n}$. Note in particular that
\begin{eqnarray} \label{basisop}
  {\mathbf{a}}^2 & = & \langle {\mathbf{a}}, {\mathbf{a}} \rangle  \notag \\
  {\mathbf{e}}_{i}^{2} & = & {\mathbf{1}} \\
  {\mathbf{e}}_{i} {\mathbf{e}}_{j} & = & -{\mathbf{e}}_{j}
  {\mathbf{e}}_{i} , \; \; \; \; i \neq j .  \notag
\end{eqnarray}

The Clifford algebra ${\mathcal{C}l}_n$ is itself a vector space of
dimension $\sum_{p=0}^{n} {\binom{n }{p}} = 2^n$, with basis
\begin{equation*}
  \left\{ 1,{\mathbf{e}}_1 , \ldots ,{\mathbf{e}}_n ,
    {\mathbf{e}}_1 {\mathbf{e}}_2, \ldots 
    ,{\mathbf{e}}_1 {\mathbf{e}}_n , \ldots 
    ,{\mathbf{e}}_1 {\mathbf{e}}_2 \cdots {\mathbf{e}}_n
  \right \},
\end{equation*}
such that an element ${\mathcal{A}}$ in ${\mathcal{C}l}_n$ is written
as
\begin{eqnarray}
 \label{multi} 
  A &=& a_0 +a_{11} {\mathbf{e}}_1 +\cdots
        +a_{1i} {\mathbf{e}}_n +a_{21} {\mathbf{e}}_1 {\mathbf{e}}_2 +
        \nonumber \\
    & & + \cdots
        a_{2i} {\mathbf{e}}_1 {\mathbf{e}}_n +\cdots +a_{di} {\mathbf{e}}_1
        {\mathbf{e}}_2 \cdots {\mathbf{e}}_n ,
\end{eqnarray}
where $d=2^{n} -1$ and $i = {\binom{n }{p}}$ for the real numbers
$a_{pi}$.  Consequently, the vector space ${\mathcal{C}l}_n$ can be
decomposed in $n+1$ subspaces as:
\begin{equation}
  \label{parti}
  {\mathcal{C}l}_n = \Lambda^0
  \mathbb{R}^n \oplus \Lambda^1 \mathbb{R}^n \oplus \cdots \oplus
  \Lambda^n \mathbb{R}^n .
\end{equation}
Each subspace is of dimension ${\binom{n }{p}}$.

The elements ${\mathcal{A}}$ (Eqn. \ref{multi}) of the Clifford
algebra ${\mathcal{C}l}_n$ are called multivectors, and those of
$\Lambda^p \mathbb{R}^n$, $p$-vectors. In particular, $0$-vectors are
real numbers and $\dim( \Lambda^0 \mathbb{R}^n ) = 1$. $\Lambda^1
\mathbb{R}^n$ has the basis $\{ {\mathbf{e}}_1 ,{\mathbf{e}}_2 ,
\ldots ,{\mathbf{e}}_n \}$, so $1$-vectors are simply vectors and
$\dim ( \Lambda^1 \mathbb{R}^n ) = n $. $\Lambda^2 \mathbb{R}^n$ has
the basis $\{ {\mathbf{e}}_1 {\mathbf{e}}_2 ,{\mathbf{e}}_1
{\mathbf{e}}_3 , \ldots ,{\mathbf{e}}_1 {\mathbf{e}}_n \}$ and their
elements ($2$-vectors) are also called bivectors. Finally, $\Lambda^n
\mathbb{R}^n$ has as basis $\{ {\mathbf{e}}_1 {\mathbf{e}}_2 \cdots
{\mathbf{e}}_n \}$ and since $\dim ( \Lambda^n \mathbb{R}^n ) = 1$,
the $n$-vectors of ${\mathcal{C}l}_n$ are referred as
pseudoscalars.

In this paper, arbitrary multivectors will be denoted by non bold
upper case characters without ornamentation such as $A$. $p$-vectors
will be denoted by $A_p$, with the exception of vectors ($1$-vectors),
that will be denoted by bold lower case characters such as
${\mathbf{a}}$.

Bearing in mind that the decomposition of the vector space
${\mathcal{C}l}_n$ as the direct sum of the subspaces $\Lambda^p
\mathbb{R}^n$, $0 \leq p \leq n$, given in (\ref{parti}), any
multivector $A$ can be written as
\begin{equation}
 \label{decomp}
 A = \langle A \rangle _{0} + \langle A \rangle _{1} + \cdots + \langle
 A\rangle _{n},
\end{equation}
where $\langle A\rangle _{p}$, the $p$-vector part of $A$, is the
projection of $A \in {\mathcal{C}l}_n$ into $\Lambda^p
\mathbb{R}^n$. $\langle \;\; \rangle$ is called the grade operator.

\subsection{Innner and outer products}
\vspace{-4pt}

Given the decomposition (\ref{parti}), an important property of a
Clifford algebra is the existence of products that allows us to move
from one subspace of ${\mathcal{C}l}_{n}$ to another. Let us first
consider the product of two $1$-vectors. For all $\mathbf{u},
\mathbf{v}\in \Lambda ^{1}\mathbb{R}^{n}$, their product $\mathbf{u}
\mathbf{v}$ can be written as
\begin{equation*}
  \mathbf{u}\mathbf{v}=\frac{1}{2} \left( \mathbf{u}\mathbf{v} + 
    \mathbf{v} \mathbf{u}\right)  + \frac{1}{2} \left( \mathbf{u} 
    \mathbf{v}  -\mathbf{v} \mathbf{u} \right) .
\end{equation*}
Now define the \textquotedblleft inner\textquotedblright\ and
\textquotedblleft outer\textquotedblright\ products as follows
\begin{eqnarray*}
  \mathbf{u}\cdot \mathbf{v} &=& \frac{1}{2} \left( \mathbf{u} 
    \mathbf{v} + \mathbf{v} \mathbf{u} \right) = \langle {\mathbf{a}}, 
  {\mathbf{b}} \rangle , \\
  \mathbf{u} \wedge \mathbf{v} &=& \frac{1}{2} \left( \mathbf{u} 
    \mathbf{v} - \mathbf{v}\mathbf{u}\right) .
\end{eqnarray*}
The inner product is symmetric and notice that vectors $\mathbf{u}$
and $\mathbf{v}$ are orthogonal if an only if $\mathbf{u} \mathbf{v}
= -\mathbf{v} \mathbf{u}$. The outer product $\mathbf{u}\wedge
\mathbf{v}$ is antisymmetric (and associative) and vanishes whenever
the two vectors are collinear, that is, $\mathbf{u}$ and $\mathbf{v}$
are collinear (or linearly dependent) if an only if
$\mathbf{u}\mathbf{v}=\mathbf{v}\mathbf{u}$. Thus, the product
$\mathbf{u}\mathbf{v}$ provides information about the relative
directions of the vectors. Anticommutativity means orthogonality and
commutativity means collinearity. Notice that for the bases vectors of
$\mathbb{R}^{n}$, we have
\begin{eqnarray*}
  \mathbf{e}_{i}\mathbf{e}_{j} &=&\mathbf{e}_{i}\wedge \mathbf{e}_{j},
  \;\;\;i\neq j, \\
  \mathbf{e}_{i}\cdot \mathbf{e}_{i} &=&\mathbf{e}_{i}^{2}=1.
\end{eqnarray*}

From the following equality 
\begin{eqnarray*}
  \mathbf{u}\mathbf{v} &=& \mathbf{u} \cdot \mathbf{v} + \mathbf{u} 
  \wedge \mathbf{v}, \\
  &=& \langle \mathbf{u} \mathbf{v} \rangle _{0} + \langle \mathbf{u} 
  \mathbf{v} \rangle _{2},
\end{eqnarray*}
we can extend the notions of inner and outer product to the case of
$p$- and $q$-vectors in the following way. For a $p$-vector $A_{p}\in
\Lambda ^{p}\mathbb{R}^{n}$ and a $q$-vector $B_{q}\in \Lambda ^{q}
\mathbb{R}^{n}$, the inner product $A_{p}\cdot B_{q}$ is defined by
\begin{equation}
 \label{innerp}
 A_p \cdot B_q = \left\{ 
\begin{array}{ll}
  \left\langle A_p B_q \right\rangle _{\left| p-q\right| } & \text{if} \quad
  p,q>0 , \\ 
  0 & \text{if} \quad p=0 \text{ or } q=0 .
\end{array}
\right.
\end{equation}

Analogously, the outer product $A_p \wedge B_q$ is defined by
\begin{equation}
  \label{outerp}
  A_p \wedge B_q = \left \langle A_p B_q \right\rangle _{p+q}.
\end{equation}

Since arbitrary multivectors can be decomposed as in (\ref{decomp}),
inner and outer product can be extended by linearity to
${\mathcal{C}l}_n$. Then, given $A, B \in {\mathcal{C}l}_n$, we have
\begin{eqnarray*}
  A \cdot B = \sum_{k,l = 1}^n \langle A \rangle_k \cdot \langle B \rangle_l,
  \\
  A \wedge B = \sum_{k,l = 1}^n \langle A \rangle_k \wedge \langle B \rangle_l,
\end{eqnarray*}

\subsection{Geometric interpretation}
\vspace{-4pt}

Bivectors have an interesting geometric interpretation. Just as a
vector describes an oriented line segment, with the direction of the
vector representing the oriented line and the magnitude of the vector, the length of the segment; a bivector ${\mathbf{a}}
\wedge {\mathbf{b}}$ describes an oriented plane segment, with the
direction of the bivector representing the oriented plane and the
magnitude of the bivector measuring the area of the plane segment
(Figure \ref{fig:fig1}). The same interpretation is extended to high-order
terms: ${\mathbf{a}} \wedge {\mathbf{b}} \wedge {\mathbf{c}}$
represents an oriented volume. In general, multivectors contain
information about orientation of subspaces.

\subsection{The main involution}
\vspace{-4pt}

The magnitude or modulus of a multivector $A$ is defined by the
equation
\begin{equation}
  \label{magnitude}
  \left| A \right| = \left\langle \widetilde{A} A\right\rangle _{0}^{1/2} ,
\end{equation}
where $\sim$ denotes the operation reverse defined as
\begin{equation*}
  \left ( {\mathbf{e}}_1 {\mathbf{e}}_2 \cdots {\mathbf{e}}_p \right )^{\sim}
  = {\mathbf{e}}_p \cdots {\mathbf{e}}_2 {\mathbf{e}}_1 .
\end{equation*}

The operation reverse is distributive \cite{HesSob} so that the
reverse of an arbitrary multivector $A$ can be easily calculated. If the inverse of a multivector $A$ exists, it is denoted by $A^{-1} $
or ${\mathbf{1}}/A$, and is defined by the equation $AA^{-1} =
{\mathbf{1}}$.

\subsection{General metrics}
\vspace{-4pt}

In many physical applications one considers real vector spaces
$\mathbb{R}^n$ with metrics that are not positive definite with the
bilinear form $\langle \; , \; \rangle $, in (\ref{product}), such
that
\begin{equation*}
  \langle {\mathbf{x}}, {\mathbf{x}} \rangle = x_{1}^{2} +x_{2}^{2} + \cdots +
  x_{p}^{2} -x_{p+1}^{2} -\cdots -x_{p+q}^{2} ,
\end{equation*}
where $n=p+q$. In this case, the vector space is denoted as
$\mathbb{R}^{p,q} $, giving rise to the Clifford algebra
${\mathcal{C}l}_{p,q}$. Using the orthonormal basis $\{ {\mathbf{e}}_1
,{\mathbf{e}}_2 ,\ldots ,{\mathbf{e}}_n \}$ of $\mathbb{R}^{p,q}$, the
relations (\ref{basisop}) are now:
\begin{eqnarray}
  \label{nbasisop}
  {\mathbf{e}}_{i}^{2} & = & {\mathbf{1}} \;\;\;\; 1 < i \leq p  \notag \\
  {\mathbf{e}}_{i}^{2} & = & -{\mathbf{1}} \;\;\;\; p < i \leq n \\
  {\mathbf{e}}_{i} {\mathbf{e}}_{j} & = & -{\mathbf{e}}_{j} {\mathbf{e}}_{i},
  \;\;\;\; i \neq j .  \notag
\end{eqnarray}

The scalar $p$ is called the signature of the bilinear form $\langle
\; , \; \rangle$.

\section{Clifford algebra calculations with \textit{Mathematica}}
\label{S3} \vspace{-4pt}

According to Eqns. (\ref{innerp}), (\ref{outerp}) and
(\ref{magnitude}), all that we should need to manipulate
multivectors in a computer algebra program such as
\textit{Mathematica}, would be to define the two basic operations:
geometric product and grade operator. In the first case, a simple
algorithm for the computation of the geometric product between
multivectors can be devised by noticing that a general multivector
(\ref{multi}) in ${\mathcal{C}l}_{p,q}$ is formed by a linear
combination of terms in the form
\begin{equation}
  \label{master} 
  {\mathbf{e}}_{1}^{m_{1} }
  {\mathbf{e}}_{2}^{m_{2} } \cdots {\mathbf{e}}_{n}^{m_{n} },
\end{equation}
where $m_i = 1,0$, $(i=1,...,n)$. Let us call blades to multivectors
of the form (\ref{master}). The geometric product of two of these
blades is:
\begin{align}
 \label{nmaster}
  & \left ( {\mathbf{e}}_{1}^{m_{1} } {\mathbf{e}}_{2}^{m_{2} } \cdots 
    {\mathbf{e}}_{n}^{m_{n} } \right ) \left ( {\mathbf{e}}_{1}^{r_{1} } 
    {\mathbf{e}}_{2}^{r_{2} } \cdots {\mathbf{e}}_{n}^{r_{n} } \right
    ) = \nonumber \\
  & \qquad (-1)^{s} {\mathbf{e}}_{1}^{m_{1} +r_{1} } {\mathbf{e}}_{2}^{m_{2} +r_{2} } 
    \cdots {\mathbf{e}}_{n}^{m_{n} +r_{n} },
\end{align}
where the sum $m_i +r_i$ is evaluated modulus two, and
\begin{equation*}
  s = \sum\limits_{1\leq i<j\leq n} r_{i} m_{j} .
\end{equation*}
If $m_i +r_i = 2$ then, in order to have the right hand side in the
form (\ref{master}) when considering the signature of the bilinear
form, ${\mathbf{e}}_{i}^{m_{i} +r_{i}}$ will be replaced with $\langle
{\mathbf{e}}_i ,{\mathbf{e}}_i \rangle {\mathbf{e}}_{i}^0$, and in
this case we have:
\begin{align*}
  &  {\mathbf{e}}_{1}^{m_{1} +r_{1} } \cdots
    {\mathbf{e}}_{i}^{m_{i} +r_{i} } \cdots {\mathbf{e}}_{n}^{m_{n}
    +r_{n} } = \\
  &\qquad \langle {\mathbf{e}}_{i} , {\mathbf{e}}_{i} \rangle
    {\mathbf{e}}_{1}^{m_{1} +r_{1} } \cdots {\mathbf{e}}_{i}^{0} \cdots
    {\mathbf{e}}_{n}^{m_{n} +r_{n} }.
\end{align*}

Equation (\ref{master}) enables us to establish an isomorphism between
blades and n-tuples $(m_{1} ,m_{2} ,...,m_{n} )$ that can be manipulated more
easily from a computational point of view. The grade of a
blade such as (\ref{master}) is simply $m_1 +m_2 + \cdots +m_n $.

If, in \textit{Mathematica} code, we denote the $j$-th basis vector
${\mathbf{e}}_{j} $ as {\small \verb|e[j]|}, a blade such as ${\mathbf{e}}_1
{\mathbf{e}}_3 {\mathbf{e}}_4 $, ( ${\mathbf{e}}_{1}^{1}
{\mathbf{e}}_{2}^{0} {\mathbf{e}}_{3}^{1} {\mathbf{e}}_{4}^{1} $ using
the nomenclature of Eqn. \ref{master}) is written as:
\begin{equation}
 \label{eq:basis}
 \verb|e[1]e[3]e[4]| ,
\end{equation}
and can be internally represented simply by $(1,0,1,1)$. Care must be
taken in preserving the canonical order of the expression, since for
instance {\small \verb|e[1]e[3]|} is a geometric product of two
vectors and {\small \verb|e[1]e[3]|} $\neq $ {\small \verb|e[3]e[1]|}.

Let us consider the Clifford algebra ${\mathcal{C}l}_{p,q}$. If
{\small \verb|dim|} $= p + q$ is the dimension of the vector space,
the following \textit{Mathematica} code implements the transformation
of a blade onto a $n$-tuple: 
{\footnotesize
\begin{verbatim}
ntuple[x_, dim_]:= 
     ReplacePart[Table[0,{dim}],1,List @@
                           x /. e[k_]->{k}]
\end{verbatim}
} The signature of the bilinear form can be set by using {\small
  \verb|$SetSignature = p|}. Relations (\ref{nbasisop}) can be coded
as: 
{\footnotesize
\begin{verbatim}
bilinearform[e[i_],e[i_]]:= 
        If[i <= $SetSignature, 1, -1]
\end{verbatim}
}
With some exceptions, it is not necessary to define the dimension of
the vector space since it can be calculated directly. The maximum
dimension of the space where a blade is embedded can be extracted from
the list:
{\footnotesize
\begin{verbatim}
dimensions[x_]:= 
    List @@ x /. e[k_?Positive] -> k
\end{verbatim}
}
and, therefore, enables us to perform computations in any dimension with any given signature. For a general multivector, we need the relations that include the
distributivity of addition:
{\footnotesize
\begin{verbatim}
dimensions[x_Plus]:= 
   List @@ Distribute[tmp[x]] /. 
                        tmp -> dimensions
dimensions[a_]:= 
   {0} /; FreeQ[a,e[_?Positive]]
dimensions[a_ x_]:= 
   dimensions[x] /; FreeQ[a,e[_?Positive]]
\end{verbatim}
} 
From (\ref{nmaster}), the geometric product between two blades with a
bilinear form of signature {\small \verb|p|} can therefore be
evaluated with {\small \verb|geoprod|}:
{\footnotesize
\begin{verbatim}
geoprod[x_,y_]:= Module[{q=1,s,r={}},
  p1= ntuple[x, Max[dimensions[x],
               dimensions[y]]],
  p2= ntuple[y, Max[dimensions[x],
               dimensions[y]]],
  s= Sum[p2[[m]]*p1[[n]], 
   {m,Length[p1]-1},{n,m+1,Length[p2]}];
  r1=p1+p2;
  r=Mod[p1+p2,2];
  Do[
    If[r1[[i]]==2,
     q*=bilinearform[e[i],e[i]]];
    If[r[[i]]==1, q*=e[i]], 
  {i, Length[r]}];
    (-1)^s*q ]
\end{verbatim}
} 
The latter function evaluates the geometric product of two blades in a
Clifford algebra ${\mathcal{C}l}_{p,q}$. One step further consists in
to calculate the geometric product of two arbitrary multivectors such
as (\ref{multi}). This can be achieved from {\small \verb|geoprod|} by
providing the transformation rules which contains the properties of
the geometric product under multiplication of blades by real numbers
and addition of blades. Here is the behavior under scalar
multiplication:
{\footnotesize
\begin{verbatim}
geoprod[a_,y_]:= 
   a y /; FreeQ[a,e[_?Positive]]
geoprod[x_,a_]:= 
   a x /; FreeQ[a,e[_?Positive]]
geoprod[a_ x_,y_]:= 
   a geoprod[x,y] /; FreeQ[a,e[_?Positive]]
geoprod[x_,a_ y_]:= 
   a geoprod[x,y] /; FreeQ[a,e[_?Positive]]
\end{verbatim}
}
\noindent and the distributivity of addition:
{\footnotesize
\begin{verbatim}
geoprod[x_, y_Plus]:= 
   Distribute[tmp[x,y,p],Plus] /. 
                        tmp -> geoprod
geoprod[x_Plus, y_]:= 
   Distribute[tmp [x,y,p],Plus] /. 
                        tmp -> geoprod
\end{verbatim}
}
Therefore, the function to calculate the geometric product of
arbitrary multivectors is defined as
{\footnotesize
\begin{verbatim}
GeometricProduct[_]:= $Failed
GeometricProduct[m1_,m2_,m3__]:= 
    tmp[ GeometricProduct[m1,m2],m3] /.
                tmp -> GeometricProduct
GeometricProduct[m1_,m2_]:= 
    geoprod[Expand[m1], Expand[m2]]
\end{verbatim}
}
To complete the basic operations of a Clifford algebra, we must define and 
implement the grade operator. The following auxiliary function
calculates the grade of a blade
{\footnotesize
\begin{verbatim}
gradblade[a_]:= 
     0 /; FreeQ[a,e[_?Positive]]
gradblade[x_]:= 
     Plus @@ ntuple[x, Max[dimensions[x]]]
gradblade[a_ x_]:= 
     gradblade[x] /; FreeQ[a,e[_?Positive]]
\end{verbatim}
}
Now, {\small \verb|Grade[x,n]|} should extract the term of grade
$n$ from the multivector $X$.
Firstly, we consider the case when the multivector $X$
is a blade of grade $r$:
$\langle X \rangle _n = 0$ if $r \neq n$ and $\langle X \rangle _n =
X$ if $r = n$. The code reads:
{\footnotesize
\begin{verbatim}
Grade[x_, n_?NumberQ]:= 
                  If[gradblade[x]==n,x,0]]
\end{verbatim}
}
For a general multivector $X$, we have:
{\footnotesize
\begin{verbatim}
Grade[x_Plus, n_?NumberQ]:= 
  Distribute[tmp[x,n],Plus] /. tmp -> Grade
\end{verbatim}
}
Functions {\small \verb|GeometricProduct|} and {\small \verb|Grade|}
enable us to construct all the operations which can be defined in a
Clifford algebra, such as outer product ({\small
  \verb|OuterProduct[v,w,..]|}), inner product ({\small
  \verb|InnerProduct[v,w,..]|}), magnitude ({\small
  \verb|Magnitude[v]|}), reverse ({\small \verb|Turn[v]}|), inverse
({\small \verb|MultivectorInverse[v]|}), dual ({\small
  \verb|Dual[v,dim]|}), and many others, all included in the package
{\small \verb|Clifford.m|}. This package works with general
multivectors of the form (\ref{multi}), but particular cases can help
to envisage the power of multivector calculus, as it is made explicit in
what follows.

\section{Vectors}
\label{S4} \vspace{-4pt}

Let us consider the Clifford algebra
${\mathcal{C}l}_{p,0}$. $n$-dimensional vectors are $1$-vectors and
lie in the subspace $\Lambda ^{1} \mathbb{R}^{p,0}$. A vector
${\mathbf{a}}$ is therefore ${\mathbf{a}} = \langle A \rangle _{1}$,
where $A$ is a general multivector.

The inner product defined in (\ref{innerp}) becomes now the standard
\textquotedblleft dot product\textquotedblright~ between vectors. The
\textquotedblleft cross product\textquotedblright~
${\mathbf{a}} \times {\mathbf{b}}$ of vector calculus is defined in
$\mathbb{R}^3$ and is related to the outer product of vectors
${\mathbf{a}}$ and ${\mathbf{b}}$. The vector
${\mathbf{a}} \times {\mathbf{b}}$ is perpendicular to
${\mathbf{a}} \wedge {\mathbf{b}}$ and with the same magnitude:
$\left| {\mathbf{a}} \wedge {\mathbf{b}} \right| = \left| {\mathbf{a}}
  \times {\mathbf{b}} \right|$.
The explicit algebraic relation between them is \cite{Hes87}:
\begin{equation}
  \label{cross}
  {\mathbf{a}} \times {\mathbf{b}} = (-{\mathbf{e}}_1 {\mathbf{e}}_2 
  {\mathbf{e}}_3 )({\mathbf{a}} \wedge {\mathbf{b}}),
\end{equation}
where ${\mathbf{e}}_1 {\mathbf{e}}_2 {\mathbf{e}}_3$ is the
pseudoscalar of ${\mathcal{C}l}_{3,0}$. We may actually take this as a
definition of the cross product.

From (\ref{cross}), it is easy to define the function {\small
  \verb|crossprod[v,w]|} that gives the cross product between two
three-dimensional vectors ${\mathbf{v}}$ and ${\mathbf{w}}$:
{\footnotesize
\begin{verbatim}
crossprod[v_,w_]:= 
      GeometricProduct[-e[1]e[2]e[3],
                         OuterProduct[v,w]]
\end{verbatim}
}
The associativity property of the geometric product allows algebraic
manipulations typical of real numbers that are not possible in the
Gibbs' vector algebra since the cross and dot products are not
generally associative. For example
\begin{equation*}
  {\mathbf{a}} \times ( {\mathbf{b}} \times
  {\mathbf{c}}) \neq ({\mathbf{a}} \times {\mathbf{b}})\times
  {\mathbf{c}}.
\end{equation*}

Even more, many products such as ${\mathbf{a}} \cdot ({\mathbf{b}}
\cdot {\mathbf{c}})$ are not even defined. With Clifford algebra all
products are not only well defined but associative making simpler many
algebraic manipulations and allowing to define derivatives and
integrals just as they are defined for real functions of real
variables, provided that we are careful to maintain the order of the
factors since geometric product is not commutative.

One specific example that shows the simplicity of some expressions if
the geometric product is used, is concerning rotations. Consider a vector
${\mathbf{v}}$ in $\mathbb{R}^n$ which is rotated by an angle $\theta$
\textit{in} the oriented plane characterized by the bivector
${\mathbf{a}} \wedge {\mathbf{b}}$. After the rotation, the vector
${\mathbf{v}}$ is transformed into ${\mathbf{v}}^{\prime}$, given by
\cite{Hes87}:
\begin{equation}
  \label{rotacion}
  {\mathbf{v}} = \widetilde{U} {\mathbf{v}} U,
\end{equation}
where
\begin{equation*}
  U = \cos (\theta /2) + \frac{{\mathbf{a}} \wedge 
    {\mathbf{b}}}{\left| {\mathbf{a}} \wedge {\mathbf{b}} \right| } 
  \sin (\theta /2).
\end{equation*}
The direction of the rotation (clockwise or counter clockwise) is
specified by the orientation of the bivector ${\mathbf{a}} \wedge
{\mathbf{b}}$. Eqn. \ref{rotacion} gives the rotated vector
${\mathbf{v}}^{\prime}$ regardless of the dimension of the space in
which it is embedded. No corresponding simple expression exists in
vector algebra.

As an example, which can be easily visualized, consider
the vector ${\mathbf{v}} = (1,1,1)$, to be rotated $90^o$ maintaining
invariant the plane $xy$. To characterize the plane $xy$, we can use
${\mathbf{a}} \wedge {\mathbf{b}}$ where ${\mathbf{a}} = (1,0,0)$ and
${\mathbf{b}} = (0,1,0)$, in which case we get a rotation
counterclockwise and is easy to see that ${\mathbf{v}}^{\prime}=
(-1,1,1)$, but if we use ${\mathbf{b}} \wedge {\mathbf{a}}$ then the
rotation is clockwise and ${\mathbf{v}}^{\prime} = (1,-1,1)$. Here is
this example solved with {\small \verb|Clifford|}:
{\footnotesize
\begin{verbatim}
In[1]:= << Clifford.m
In[2]:= v = e[1]+e[2]+e[3];
In[3]:= plane = OuterProduct[e[1],e[2]]/
         Magnitude[OuterProduct[e[1],e[2]]]
Out[3]= e[1]e[2]
\end{verbatim}
}
The operator $U$ is now defined ($\widetilde{U}$ is the reverse of
$U$):
{\footnotesize
\begin{verbatim}
In[4]:= u = Cos[Pi/4] + plane Sin[Pi/4];
In[5]:= vprime = GeometricProduct[Turn[u],
                     GeometricProduct[v,u]]
Out[5]= -e[1]+e[2]+e[3]
\end{verbatim}
} 
So we get a counterclockwise rotation where the vector
${\mathbf{v}}=(1,1,1)$ becomes ${\mathbf{v}}^{\prime }=(-1,1,1)$. The
implementation in \textit{Mathematica} of this rotation is the
function {\small \verb|Rotation[v,a,b]|}.

\section{Drawing multivectors in the 3-dimensional space.}
\label{S5} \vspace{-4pt}

The elements of the Clifford algebra ${\mathcal{C}l}_{3,0}$ can be
visualized in the $3$-dimensional space, providing geometrical
insights. The package include a function to draw multivectors
belonging to ${\mathcal{C}l}_{3,0}$, called \verb|GADraw|, that will
be described in what follows.

{\small \verb|GADraw|} includes embedded functions to draw vectors,
bivectors and the pseudoscalar of ${\mathcal{C}l}_{3,0}$. For example
let us draw the multivector
${\mathbf{e}}_{3} + {\mathbf{e}}_{1}{\mathbf{e}}_{2} +
{\mathbf{e}}_{2}{\mathbf{e}}_{3} + {\mathbf{e}}_{1}{\mathbf{e}}_{2}
{\mathbf{e}}_{3}$:
{\footnotesize
\begin{verbatim}
In[1]:= << Clifford.m
In[2]:= A = e[3] + e[1]e[2] +
                     e[2]e[3]+e[1]e[2]e[3];
In[3]:= GADraw[A];
Out[3]= 
\end{verbatim}
}

\begin{figure}[!h]
\begin{center}
\includegraphics[width=8.0cm]{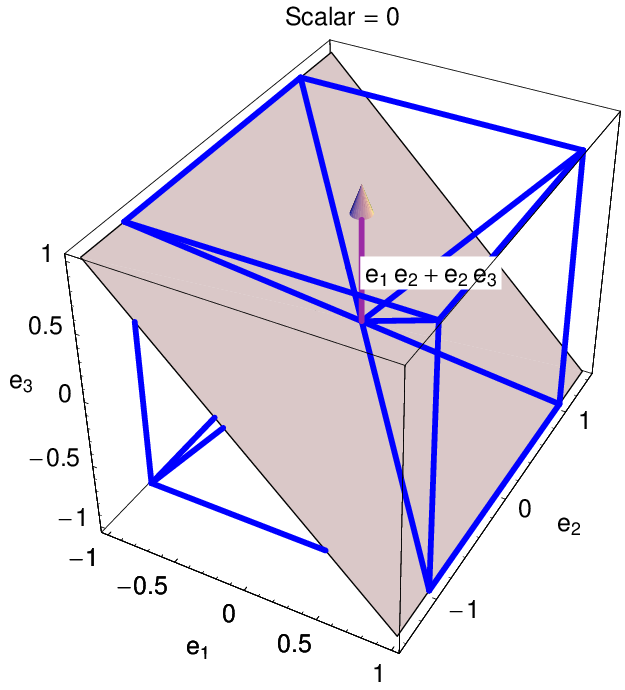}
\end{center}
\caption{A vector, a bivector (plane) and the pseudoscalar (cube)
  drawn with the aid of the function \texttt{GADraw}.}
\label{fig:fig1}
\end{figure}

The result is shown in Fig. \ref{fig:fig1}. The bivector is
represented by an area and the pseudoscalar as a scalable cube. In the
particular case of the vector $(0,0,1)$, the arrow in its tip was
generated by the following code \cite{MathGraph}:
{\footnotesize
\begin{verbatim}
mat[1] = Sin[t]*(e[1]/14)+Cos[t]*(e[2]/14), 
mat[2] = Sin[t+0.25]*(e[1]/14) + 
                     Cos[t+0.25]*(e[2]/14),
mat[3] = e[3]/5,
\end{verbatim}
}
This arrow is then translated and rotated (in this case to the tip of
the vector $(0,0,1)$) with the aid of the function {\small
  \verb|Rotation[mat,w,p]|}, where $w = e_{3}$ and $p$ is the vector
that points the site where the tip of the arrow must be located. If
{\small \verb|sc=Sqrt[p[[1]]^2 +p[[2]]^2 +p[[3]]^2]/2|} is a scale
factor then all this procedure can be encoded as:
{\footnotesize
\begin{verbatim}
If[OuterProduct[ToBasis[p], e[3]] === 0,
cone =
 Table[Array[ToVector[mat[#],3] &,3] +
   p-ToVector[mat[3],3],{t,0.25,2*Pi,0.25}],
elms = 
 Array[sc*ToVector[Grade[Rotation[
         mat[#],e[3], ToBasis[p]],1],3]&,3];
cone=
 Table[elms+p-res[[3]],{t,0.25,2*Pi,0.25}]];

 arrow = Graphics3D[{FaceForm[color], 
              EdgeForm[], Polygon /@ cone}]
\end{verbatim}
}
Where {\small \verb|ToBasis|} and {\small \verb|ToVector|} are
functions, defined on the same package {\small \verb|Clifford|}; the
former changes from the coordinates of a vector to the basis notation
(\ref{eq:basis}) and the last one does the opposite.

We must emphasize that the above code was included with the aim
to show that symmetry operations, rotations in this case, can be
carried out by using the elements of the Clifford algebra without
requiring matrices. This algebra provides a consistent computational
framework with significant applications in computer graphics, vision
and robotics \cite{Dorst}.

\section{Complex numbers}
\label{S6} \vspace{-4pt}

Let us consider the Clifford algebra of the most simple space that has
a geometrical structure: the plane $\mathbb{R}^{2,0}$. Taking the
canonical basis $\{ {\mathbf{e}}_1 ,{\mathbf{e}}_2 \}$, a basis for
the Clifford algebra ${\mathcal{C}l}_{2,0}$ is $\{ {\mathbf{1}},
{\mathbf{e}}_1 ,{\mathbf{e}}_2 , {\mathbf{e}}_1 {\mathbf{e}}_2 \}$ and
a general multivector $A \in {\mathcal{C}l}_{2,0}$ has the form
\begin{equation*}
  A = k_0 +k_1 {\mathbf{e}}_1 +k_2 {\mathbf{e}}_2 +k_3 {\mathbf{e}}_1 
  {\mathbf{e}}_2 .
\end{equation*}

We can decompose ${\mathcal{C}l}_{2,0}$ as ${\mathcal{C}l}_{2,0} =
{\mathcal{C}l}_{2,0}^{+} \oplus {\mathcal{C}l}_{2,0}^{-}$, such that
${\mathcal{C}l}_{2,0}^{+} $ contains even grade elements and
${\mathcal{C}l}_{2,0}^{-} $ contains odd grade elements. Therefore,
$A$ can be expressed as the sum of two multivectors: $A = A^{+} +
A^{-}$, where $A^{+} \in {\mathcal{C}l}_{2,0}^{+}$ and $A^{-} \in
{\mathcal{C}l}_{2,0}^{-}$. That is
\begin{eqnarray}
  A & = & A^{+} +A^{-},  \notag \\
  A^{+} & = & k_{0} +k_{3} {\mathbf{e}}_{1} {\mathbf{e}}_{2},  \notag \\
  A^{-} & = & k_{1} {\mathbf{e}}_{1} +k_{2} {\mathbf{e}}_{2}.  \notag
\end{eqnarray}

We focus our attention into ${\mathcal{C}l}_{2,0}^{+}$; it is itself
an algebra so that it is called the even subalgebra of
${\mathcal{C}l}_{2,0}$.  By taking
\begin{equation}
  \label{complex}
  {\mathbf{i}} = {\mathbf{e}}_1 {\mathbf{e}}_2 ,
\end{equation}
an element of ${\mathcal{C}l}_{2,0}^{+}$ can be written as
\begin{equation*}
z = k_{0} +k_{3} {\mathbf{i}}.
\end{equation*}
Since ${\mathbf{i}}^{2} =({\mathbf{e}}_1 {\mathbf{e}}_2
)({\mathbf{e}}_1 {\mathbf{e}}_2 ) = -({\mathbf{e}}_1 {\mathbf{e}}_2
{\mathbf{e}}_2 {\mathbf{e}}_1 ) = -1$, and ${\mathbf{i}}$ is itself a
generator of rotations \cite{Hes87}, we see that
${\mathcal{C}l}_{2,0}^{+}$ is equivalent to the algebra of complex
numbers.

The algebraic operations of complex numbers can therefore be worked
out with {\small \verb|Clifford|}. In order to get a more standard notation we
firstly define the function
{\footnotesize
\begin{verbatim}
In[6]:= transform[x]:= x /. i -> e[1]e[2]
\end{verbatim}
}
which makes the identification (\ref{complex}). Here we have two
complex numbers:
{\footnotesize
\begin{verbatim}
In[7]:= w = a + b i;
In[8]:= z = c + d i;
\end{verbatim}
}
The product $(a + b {\mathbf{i}})(c + d {\mathbf{i}})$ becomes
{\footnotesize
\begin{verbatim}
In[9]:= GeometricProduct[transform[w],
             transform[z]] /. e[1]e[2] -> i
Out[9]:= a c - b c + b c i + a d i
\end{verbatim}
}
The equivalence between some built-in basic operations of complex
numbers in \textit{Mathematica} and those of a Clifford algebra
defined the package is shown in the following table:

\begin{table}[h!]
 \begin{center}
\begin{tabular}{ll}
  Built-in Objects & \verb|Clifford.m| Objects \\ 
                   &  \\ 
  {\small \verb|Re[z]|} & {\small \verb|Grade[z,0]|} \\ 
  {\small \verb|Im[z]|} & {\small \verb|GeometricProduct[|}
  \\ 
         &  {\small \verb|  Grade[z,2],-e[1]e[2]]|} \\
  {\small \verb|Conjugate[z]|} & {\small \verb|Turn[z]|} \\ 
  {\small \verb|Abs[z]|} & {\small \verb|Magnitude[z]|} \\ 
                   & 
\end{tabular}
\end{center}
\end{table}
The inverse of the complex number $w$ can be calculated with
{\small \verb|MultivectorInverse|}:
{\footnotesize
\begin{verbatim}
In[10]:= MultivectorInverse[transform[w]] 
                           /. e[1]e[2] -> i
 
           a - bi
Out[10]= ----------
            2    2
           a  + b
\end{verbatim}
}

\section{Quaternions}
\label{S7} \vspace{-4pt}

Extending from the previous section, we now develop
the Clifford algebra of $\mathbb{R}^{3,0}$, equipped with the
canonical basis $\{ {\mathbf{e}}_1 ,{\mathbf{e}}_2 ,{\mathbf{e}}_3
\}$. The Clifford algebra ${\mathcal{C}l}_{3,0}$ is an
eighth-dimensional vector space with the basis $\{ {\mathbf{1}},
{\mathbf{e}}_1 ,{\mathbf{e}}_2 ,{\mathbf{e}}_3 , {\mathbf{e}}_1
{\mathbf{e}}_2 ,{\mathbf{e}}_2 {\mathbf{e}}_3 ,{\mathbf{e}}_1
{\mathbf{e}}_3 ,{\mathbf{e}}_1 {\mathbf{e}}_2 {\mathbf{e}}_3 \}$. A
general multivector $A \in {\mathcal{C}l}_{3,0}$ is written as
\begin{eqnarray*}
  A &=& k_0 +k_1 {\mathbf{e}}_1 +k_2 {\mathbf{e}}_2 +k_3 {\mathbf{e}}_3 
        + k_4 {\mathbf{e}}_1 {\mathbf{e}}_2 + \\
    & & k_5 {\mathbf{e}}_2 {\mathbf{e}}_3 + k_6 {\mathbf{e}}_1 {\mathbf{e}}_3 +
        k_7 {\mathbf{e}}_1 {\mathbf{e}}_2  {\mathbf{e}}_3 ,
\end{eqnarray*}
which can be also expressed as $A = A^{+} + A^{-}$, where
\begin{eqnarray}
  A^{+} & = & k_0 +k_4 {\mathbf{e}}_1 {\mathbf{e}}_2 + k_5 {\mathbf{e}}_2 
  {\mathbf{e}}_3 +k_6 {\mathbf{e}}_1 {\mathbf{e}}_3  \notag \\
  A^{-} & = & k_1 {\mathbf{e}}_1 +k_2 {\mathbf{e}}_2 +k_3 {\mathbf{e}}_3 
  + k_7   {\mathbf{e}}_1 {\mathbf{e}}_2 {\mathbf{e}}_3 .  \notag
\end{eqnarray}

The even-grade elements $A^{+}$ form the subalgebra
${\mathcal{C}l}_{3,0}^{+} $ of ${\mathcal{C}l}_{3,0}$, equivalent to
the algebra of quaternions. Hence, we have:
\begin{eqnarray}
  \label{qide}
  {\mathbf{i}} & = & -{\mathbf{e}}_2 {\mathbf{e}}_3,  \notag \\
  {\mathbf{j}} & = & {\mathbf{e}}_1 {\mathbf{e}}_3 \\
  {\mathbf{k}} & = & -{\mathbf{e}}_1 {\mathbf{e}}_2 ,  \notag
\end{eqnarray}
leading to the famous equations
\begin{eqnarray}
  \label{qprop}
  {\mathbf{i}}^{2} & = & {\mathbf{j}}^{2} = {\mathbf{k}}^{2} = -1, \\
  \mathbf{ijk} & = &-1.  \notag
\end{eqnarray}

With the identifications given in (\ref{qide}), an element of
${\mathcal{C}l}_{3,0}^{+}$ can be written now as
\begin{equation*}
  Q = q_0 +q_1 {\mathbf{i}} +q_2 {\mathbf{j}} +q_3 {\mathbf{k}},
\end{equation*}
which, in view of the properties (\ref{qprop}), is a quaternion.

The algebra of quaternions is therefore comprised in the same
package. The basic operations of this algebra are carried out by the
function already defined, such as {\small \verb|GeometricProduct|},
{\small \verb|MultivectorInverse|}, {\small \verb|Magnitude| and
  \verb|Turn|}. To simplify the operations, we have incorporated the
definitions (\ref{qide}) and redefined some functions to work only
with quaternions and complex numbers. The new functions begin with the
word {\small \verb|Quaternion|}, namely, {\small
  \verb|QuaternionProduct|}, {\small \verb|QuaternionInverse|},
{\small \verb|QuaternionMagnitude|} and {\small
  \verb|QuaternionTurn|}. So, for instance the inverse of the
quaternion $q = a + 3 {\mathbf{i}} + 6 {\mathbf{j}} - 10 {\mathbf{k}}$
is {\footnotesize
\begin{verbatim}
In[11]:= q = a + 3 i + 6 j - 10 k;
In[12]:= QuaternionInverse[q]
 
         a - 3 i - 6 j + 10 k
Out[12]= ---------------------
                     2
              145 + a
\end{verbatim}
}
\section{Grassmann algebra}
\label{S8} \vspace{-4pt}

The outer product ({\small \verb|OuterProduct|}) defined in
(\ref{outerp}) is associative and the identity ${\mathbf{1}}$, in a
Clifford algebra ${\mathcal{C}l}_{n,0}$, is also the identity for the
outer product.  Consequently, the vector space $\mathbb{R}^{p,0}$ with
the outer product already defined is an algebra of dimension $2^n$,
which is called the \textit{Grassmann algebra} of
$\mathbb{R}^{p,0}$. Notice that it does not depend on the inner
product of the vector space, but just on the alternation of the outer
product.

Grassmann algebra has relevance in modern theoretical physics
\cite{Lasenby} and has the right structure for the theory of
determinants. The structure of this algebra contained in
Clifford algebra allows us to reformulate Grassmann calculus
\cite{Lasenby}, and to give an extensive treatment of determinants
\cite{HesSob}, both in terms of Clifford algebra.

\section{The hyperbolic plane}
\label{S9} \vspace{-4pt}

Perhaps the simplest example of a problem involving non positive
definite metrics is the Minkowsky model of the hyperbolic plane. Here
we shall develop some basic ideas and calculations just to give a
flavor of this kind of application of Clifford algebra and the use of
the package {\small \verb|Clifford|} for non positive definite
metrics. In the next section, one extra dimension is introduced and
the resulting algebras have great relevance in relativity theory and
quantum mechanics.

One can always visualize a surface of constant positive Gaussian
curvature $1 /R^2$ (see for instance \cite{Struik} for concepts
related to differential geometry) as a sphere, with radius $R$,
embedded in a three dimensional Euclidean space. The surface is
described by the equation $x_{1}^{2} +x_{2}^{2} +x_{3}^{2} = R^{2}$. A
surface of constant \textit{negative} curvature, however, cannot be
embedded in a Euclidean space, so alternative possibilities must be
developed to visualize such surfaces. The simplest surface of constant
negative curvature is often called the \textit{hyperbolic plane}, the
\textit{Bolyai-Lobachevsky plane}, or the \textit{pseudosphere}. This
surface can be globally embedded in a space equipped with the
Minkowsky metrics instead of the Euclidean one. A three-dimensional
Minkowsky space can be identified by the fact that if $(x_1 ,x_2 , x_3
)$ are the coordinates of a vector ${\mathbf{x}}$ in this space, then
the distance to the origin is $\left| x\right| ^{2} = x_{1}^{2}
+x_{2}^{2} -x_{3}^{2}$.

The equation
\begin{equation}
 \label{hyper}
 x_{1}^{2} +x_{2}^{2} -x_{3}^{2} = -R^{2} ,
\end{equation}
defines a hyperboloid of two sheets intersecting the $x_3$ axis at the
points $\pm 1$. Either sheet (upper or lower) models an infinite
surface without a boundary (the Minkowsky metric becomes positive
definite upon it) that, as we shall see, has constant Gaussian
curvature $-1 /R^2$.

We can easily convince ourselves that ${\mathcal{C}l}_{2,1}$ is indeed
the three-dimensional Minkowsky space (the signature of the bilinear
form is $2$). We shall proceed to calculate the Gaussian curvature of
the hyperboloid (\ref{hyper}) with the standard formulas of
differential geometry but with the metrics of
${\mathcal{C}l}_{2,1}$. In three dimensions, the Gaussian curvature of
a surface $f( x_1 ,x_2 ,x_3 )=0$ can be written as \cite{Weatherburn}:
\begin{equation*}
  k=\frac{1}{2} \left[ {\mathbf{n}} \cdot \nabla ^{2} 
    {\mathbf{n}} + \left( \nabla \cdot {\mathbf{n}} \right)^{2} \right] ,
\end{equation*}
where ${\mathbf{n}} = \nabla f(x_{1} ,x_{2} ,x_{3} ) / \left| \nabla
  f(x_{1} ,x_{2} ,x_{3} ) \right| $ is the normal to the surface. In a
three-dimensional space with metrics non positive definite, the
gradient of a scalar function $\phi$ and the divergence and Laplacian
of a vector function ${\mathbf{f}} = (f_1 ,f_2 ,f_3 )$ are defined as
\begin{eqnarray*}
  \nabla \phi & = & ( {\mathbf{e}}_1 \cdot {\mathbf{e}}_1 ) \frac{\partial
                    \phi }{\partial x_1 } {\mathbf{e}}_1 + ( {\mathbf{e}}_2 \cdot {\mathbf{e}}_2
                    ) \frac{\partial \phi} {\partial x_2 } {\mathbf{e}}_2 + \\
              & & ({\mathbf{e}}_3 \cdot {\mathbf{e}}_3 ) \frac{\partial \phi
                  }{\partial x_3 } {\mathbf{e}}_3 , \\
  \nabla \cdot {\mathbf{f}} & = & ( {\mathbf{e}}_1 \cdot {\mathbf{e}}_1 )^{2} 
                                  \frac{\partial f_1 }{\partial x_1 } + ( {\mathbf{e}}_2 \cdot {\mathbf{e}}_2
                                  )^{2} \frac{\partial f_2 }{\partial x_2 } + \\
              &  &  ( {\mathbf{e}}_3 \cdot  {\mathbf{e}}_3 )^{2} \frac{\partial
                   f_3 }{\partial x_3 } ,  \\ 
  \nabla ^{2} {\mathbf{f}} & = & \left[ ( {\mathbf{e}}_1 \cdot {\mathbf{e}}_1
                                 )^{3} \frac{\partial ^{2}}{\partial x_{1}^{2} } +({\mathbf{e}}_2 \cdot 
                                 {\mathbf{e}}_2 )^{3} \frac{\partial
                                 ^{2}}{\partial x_{2}^{3} } + \right . \\
         & & \left . ({\mathbf{e}}_3 \cdot {\mathbf{e}}_3 )^{3} 
                                 \frac{\partial ^{2}}{\partial x_{3}^{2} } \right] 
                                 \left( f_{1} {\mathbf{e}}_1 + f_{2} {\mathbf{e}}_{2} + 
                                 f_{3} {\mathbf{e}}_3 \right) . 
\end{eqnarray*}

We can use {\small \verb|Clifford|} to evaluate all these expressions:
{\footnotesize
\begin{verbatim}
In[1]:= << Clifford.m
\end{verbatim}
}
The adequate metrics is defined
{\footnotesize
\begin{verbatim}
In[2]:= $SetSignature = 2
\end{verbatim}
}
Here are the differential operators:
{\footnotesize
\begin{verbatim}
In[3]:= var = {x1,x2,x3};
In[4]:= GeoGrad[g_]:= 
         Sum[InnerProduct[e[k],e[k]]
           *D[g,var[[k]]]*e[k],{k,3}]
In[5]:= GeoDiv[v_]:= 
         Sum[InnerProduct[e[k],e[k]]^2 
           D[Coeff[v,e[k]],var[[k]]],{k,3}]
In[6]:= GeoLap[v_]:= 
         Sum[(InnerProduct[e[k],e[k]]^3)*
           D[v,{var[[k]],2}],{k,3}]
\end{verbatim}
}
The function {\small \verb|Coeff[m,b]|}, extracts the coefficient of
the blade {\small \verb|b|} in the multivector {\small \verb|m|}. The
surface and their normal are:
{\footnotesize
\begin{verbatim}
In[7]:= f := x1^2 + x2^2 - x3^2 + R^2;
In[8]:= norm = GeoGrad[f] / 
                     Magnitude[GeoGrad[f]];
\end{verbatim}
}
and, finally, the Gaussian curvature is:
{\footnotesize
\begin{verbatim}
In[9]:= kGauss = 
          (InnerProduct[norm,GeoLap[norm]]+
           (GeoDiv[norm])^2)/2  // Simplify
 
                 1
Out[9]= --------------------
            2     2     2
          x1  + x2  - x3
\end{verbatim}
}
Points in the surface fulfills:
{\footnotesize
\begin{verbatim}
In[10]:= x3 = Sqrt[x1^2 + x2^2 + R^2],
\end{verbatim}
}
and therefore:
{\footnotesize
\begin{verbatim}
In[11]:= kGauss
 
           -2
Out[11]= -R
\end{verbatim}
}
which is the Gaussian curvature of the hyperboloid.

Some considerations concerning the isometries (distance-preserving
transformations) of the hyperbolic plane are pertinent before leaving
this section. Like the Euclidean ones, the isometries of the
hyperbolic plane can be described in terms of reflections about given
axes. All these isometries have simple expressions in terms of
Clifford algebra. For instance, Eqn. (\ref{rotacion}) for rotations in a
given plane remains valid in ${\mathcal{C}l}_{2,1}$ but now is called a
Lorentz transformation. In general, given the vectors ${\mathbf{u}}_1
,{\mathbf{u}}_2 ,\ldots ,{\mathbf{u}}_k $ in ${\mathcal{C}l}_{p,q}$,
such that $({\mathbf{u}}_1 ,{\mathbf{u}}_2 ,\ldots , {\mathbf{u}}_k )
({\mathbf{u}}_k ,{\mathbf{u}}_{k-1} ,\ldots ,{\mathbf{u}}_1) = 1$, the
transformation
\begin{eqnarray*}
  {\mathbf{v}} &\longmapsto & ( -1)^k ( {\mathbf{u}}_k
  {\mathbf{u}}_{k-1} \cdots {\mathbf{u}}_1 ) {\mathbf{v}} (
  {\mathbf{u}}_1 {\mathbf{u}}_2 \cdots {\mathbf{u}}_k )  \\
  & &  = ( -1)^k \widetilde{U} {\mathbf{v}} U ,
\end{eqnarray*}
is an isometry \cite{HesSob}. If $k$ is even, the isometry is a
rotation, if $k$ is odd, it is a reflection \cite{Aragon2}. Now, if $q
= 0$ ($\in {\mathcal{C}l}_{p,q}$), transformations such as the
previous one are orthogonal. For $q = 1$ (as the case of the Minkoswky
space) they are Lorentz transformations.

\section{Dirac and Pauli algebras}
\label{S10} \vspace{-4pt}

Now let us add one extra dimension to the space of the previous
example and consider ${\mathcal{C}l}_{3,1}$. This vector space is
fifteen-dimensional with basis
\begin{equation*}
 \begin{aligned}
&  \left\{ \mathbf{1}, \mathbf{e}_{1} , \mathbf{e}_{2} , \mathbf{e}_{3} , 
  \mathbf{e}_{4} , \mathbf{e}_{1} \mathbf{e}_{2} , \mathbf{e}_{1} 
  \mathbf{e}_{3} , \mathbf{e}_{1} \mathbf{e}_{4} , \mathbf{e}_{2} 
  \mathbf{e}_{3} , \mathbf{e}_{2} \mathbf{e}_{4} , \mathbf{e}_{3} 
  \mathbf{e}_{4} , \right . \\
&  \left . \mathbf{e}_{1} \mathbf{e}_{2} \mathbf{e}_{3} , 
  \mathbf{e}_{1} \mathbf{e}_{3} \mathbf{e}_{4} , 
  \mathbf{e}_{2} \mathbf{e}_{3} \mathbf{e}_{4} , \mathbf{e}_{1} 
  \mathbf{e}_{2} \mathbf{e}_{3} \mathbf{e}_{4} \right \}.
 \end{aligned}
\end{equation*}

The basis vectors $\mathbf{e}_i$ satisfy the relations:
\begin{equation*}
  {\mathbf{e}}_{1}^{2} = {\mathbf{e}}_{2}^{2} =
  {\mathbf{e}}_{3}^{2} = {\mathbf{1}} , \; \; \; \;
  {\mathbf{e}}_{4}^{2} = -{\mathbf{1}},
\end{equation*}
which is the algebra of the Dirac matrices in relativistic quantum
theory.  Due to this isomorphism, ${\mathcal{C}l}_{3,1}$ is often
referred as the \textit{Dirac algebra}.

The even subalgebra ${\mathcal{C}l}_{3,1}^{+}$ has a basis $\{
{\mathbf{1}}, {\mathbf{e}}_1 {\mathbf{e}}_2 ,{\mathbf{e}}_1
{\mathbf{e}}_3 ,{\mathbf{e}}_1 {\mathbf{e}}_4 ,{\mathbf{e}}_2
{\mathbf{e}}_3 , {\mathbf{e}}_2 {\mathbf{e}}_4 , {\mathbf{e}}_3
{\mathbf{e}}_4 ,{\mathbf{e}}_1 {\mathbf{e}}_2 {\mathbf{e}}_3
{\mathbf{e}}_4 \}$ and is equivalent to the algebra of the Pauli
matrices used in quantum mechanics of spin-$\frac{1}{2}$
particles. This can be seen with the identifications
\begin{equation*}
  \sigma _{1} = {\mathbf{e}}_1 {\mathbf{e}}_4 , \; \; \; 
  \sigma _{2} = {\mathbf{e}}_2 {\mathbf{e}}_4 , \; \; \; 
  \sigma _{3} = {\mathbf{e}}_3 {\mathbf{e}}_4 .
\end{equation*}

By taking ${\mathbf{i}} = {\mathbf{e}}_1 {\mathbf{e}}_2 {\mathbf{e}}_3
{\mathbf{e}}_4 = \sigma _1 \sigma _2 \sigma _3 $, we get
\begin{eqnarray}
  \sigma _{i}^{2} & = & {\mathbf{1}}, \; \; \; i=1,2,3  \notag \\
  \sigma _{1} \sigma _{2} & = & {\mathbf{i}} \sigma _{3}  \notag \\
  \sigma _{2} \sigma _{3} & = & {\mathbf{i}} \sigma _{1}  \notag \\
  \sigma _{3} \sigma _{1} & = & {\mathbf{i}} \sigma _{2} ,  \notag
\end{eqnarray}
which are the familiar Pauli matrix
relations. ${\mathcal{C}l}_{3,1}^{+}$ is also called the \textit{Pauli
  algebra}.

Following the same reasoning, we can prove that the even subalgebra of
the Pauli algebra is isomorphic to the quaternions. The even
subalgebra of the quaternions is isomorphic to the complex
numbers. The even subalgebra of the complex numbers is $\mathbb{R}$.

\section{Conclusion}
\label{S11} \vspace{-4pt}

The basic ideas of the Clifford algebra of a vector space are
presented and a \textit{Mathematica} package for calculations within
this algebra is developed and demonstrated for complex numbers,
quaternions, the hyperbolic plane, Grassmann algebra and, Dirac and
Pauli algebras. The relevance of Clifford algebra in physics and
mathematics lies in the fact that it provides a complete algebraic
framework of geometric concepts such as directed lines, areas,
volumes, etc. (For this reason, Clifford algebra is also referred as
\textit{Geometric algebra} \cite{HesSob}). Quantities such as vectors,
complex numbers, quaternions, Pauli and Dirac matrices, have been
normally described by physicists with a mixture of disjoint
mathematical systems. All of them are naturally contained in a
Clifford algebra. It becomes therefore an efficient mathematical
language in a vast domain of physics. Future work comprises extending
our Mathematica package to other high-level programming languages in
order to facilitate its adoption within the community.

\vspace{10pt} \noindent {\bf Acknowledgements:} \ JLA wishes to thank
UNAM-DGAPA-PAPIIT for financially support this research through grant
IN106115.


\begin{thebibliography}{11}

\vspace{-7pt}
\bibitem{Ablamowicz} R.~Ab{\l}amowicz and B.~Fauser. \newblock
  Clifford/bigebra, a maple package for clifford (co)algebra
  computations, 2007.  \newblock \copyright 1996-2007,
  RA\&BF. Available at www.math.tntech.edu/rafal/.

\vspace{-7pt}
\bibitem{Aragon2} G.~Arag\'{o}n-Gonz\'{a}lez, J.L. Arag\'{o}n, and
  M.A. Rodr\'{\i}guez. The decomposition of an orthogonal
  transformation as a product of reflections.  {\em J. Math. Phys.}~
  47, 2006, \newblock Art. No. 013509 (10 pages).

\vspace{-7pt}
\bibitem{Bayro} E.~Bayro-Corrochano and G.~Sobczyk.  \newblock {\em
    Geometric Algebra with Applications in Science and Engineering}.
  \newblock Birkhauser, Boston 2001.

\vspace{-7pt}
\bibitem{Delanghe} R.~Delanghe, F.~Sommen, and V.~Soucek.  \newblock
  {\em Clifford Algebra and Spinor-Valued Functions}.  \newblock
  D. Reidel Publishing Co., Holland 1992.

\vspace{-7pt}
\bibitem{Dorst} L.~Dorst, D.~Fontijne, and S.~Mann.  \newblock {\em
    Geometric Algebra for Computer Science: An Object-Oriented
    Approach to Geometry}.  \newblock Morgan Kauffman, San Francisco,
  CA 2007.

\vspace{-7pt}
\bibitem{Ehitzer} S.~Sangwine, and E.~Hitzer.  \newblock {\em Clifford
    Multivector Toolbox: A toolbox for computing with Clifford
    algebras in Matlab}.  \newblock
  http://clifford-multivector-toolbox.sourceforge.net. Created June
  22, 1025. Last accessed December 1, 2015.

\vspace{-7pt}
\bibitem{Hes87} D.~Hestenes.  \newblock {\em New Fundations for
    Classical Mechanics}.  \newblock D. Reidel Publishing Co., Holland
  1987.

\vspace{-7pt}
\bibitem{HesSob} D.~Hestenes and G.~Sobczyk.  \newblock {\em Clifford
    Algebra to Geometric Calculus. A unified Language for Mathematics
    and Physics}.  \newblock D. Reidel Publishing Co., Holland 1985.

\vspace{-7pt}
\bibitem{HesZieg} D.~Hestenes and R.~Ziegler.  \newblock Projective
  geometry with clifford algebra.  \newblock {\em Acta Appl. Math.},
  23, 1991, 25--63.

\vspace{-7pt}
\bibitem{Jacobson} N.~Jacobson.  \newblock {\em Lectures in Abstract
    Algebra II: Linear Algebra}.  \newblock Springer, New York 1984.

\vspace{-7pt}
\bibitem{Jancewicz} B.~Jancewicz.  \newblock {\em Multivectors and
    Clifford Algebra in Electrodynamics}.  \newblock World Scientific,
  Singapore 1988.

\vspace{-7pt}
\bibitem{Kanatani} K.~Kanatani.  \newblock {\em Understanding
    Geometric Algebra: Hamilton, Grassmann, and Clifford for Computer
    Vision and Graphics}.  \newblock A K Peters/CRC Press, Florida 2015.

\vspace{-7pt}
\bibitem{Lang} S.~Lang.  \newblock {\em Linear Algebra}.  \newblock
  Addison-Wesley, New York 1969.

\vspace{-7pt}
\bibitem{Lasenby} A.~Lasenby, C.~Doran, and S.~Gull.  \newblock
  Grassmann calculus, pseudoclassical mechanics, and geometric
  algebra.  \newblock {\em J. Math. Phys.}~ 34, 1993, pp.~3683--3712.

\vspace{-7pt}
\bibitem{Lounesto}
P.~Lounesto.
\newblock {\em Clifford Algebras and Spinors}.
\newblock Cambridge University Press, Cambridge, 2001.

\vspace{-7pt}
\bibitem{Cabrera} D.~Park, R.~Cabrera, and J.-F. Gouyet.  \newblock
  Tensorial 4.0 - a mathematica package for tensor calculus.
  \newblock {\em MiER}~12, 2097, pp.~109--122.

\vspace{-7pt}
\bibitem{Porteous}
Ian Porteous.
\newblock {\em Clifford Algebras and the Classical Groups}.
\newblock Cambridge University Press, Cambridge, 1995.

\vspace{-7pt}
\bibitem{Aragon1} M.A. Rodr\'{\i}guez, J.L. Arag\'{o}n, and
  L.~Verde-Star.  \newblock Clifford algebra approach to the
  coincidence problem for planar lattices.  \newblock {\em Acta
    Crystallogr. A}~61, 2005, pp.~173--184.

\vspace{-7pt}
\bibitem{Sommer} G.~Sommer.  \newblock {\em Geometric Computing with
    Clifford Algebras Theoretical Foundations and Applications in
    Computer Vision and Robotics}.  \newblock Springer, Heidelberg
  2001.

\vspace{-7pt}
\bibitem{Struik} D.J. Struik.  \newblock {\em Differential Geometry}.
  \newblock Addison-Wesley, New York 1961.

\vspace{-7pt}
\bibitem{Weatherburn} C.E. Weatherburn.  \newblock {\em Differential
    Geometry of Three Dimensions. Vol. 1}.  \newblock Cambridge
  University Press, Cambridge 1965.

\vspace{-7pt}
\bibitem{MathGraph} T.~Wickam-Jones.  \newblock {\em Mathematica
    Graphics: Techniques \& Applications}.  \newblock Springer
  --Verlag, New York 1994.

\end{thebibliography}
\end{document}